# Magnetic oscillations driven by the spin Hall effect in 3-terminal magnetic tunnel junction devices

Luqiao Liu[1], Chi-Feng Pai[1], D. C. Ralph[1,2], R. A. Buhrman[1]

[1] Cornell University, Ithaca, NY 14853

[2] Kavli Institute at Cornell, Ithaca, NY 14853

We show that direct current in a tantalum microstrip can induce steady-state magnetic oscillations in an adjacent nanomagnet through spin torque from the spin Hall effect (SHE). The oscillations are detected electrically via a magnetic tunnel junction (MTJ) contacting the nanomagnet. The oscillation frequency can be controlled using the MTJ bias to tune the magnetic anisotropy. In this 3-terminal device the SHE torque and the MTJ bias therefore provide independent controls of the oscillation amplitude and frequency, enabling new approaches for developing tunable spin torque nano-oscillators.

A spin current can be used to excite persistent oscillations of a nanomagnet [1-2] through the spin torque (ST) mechanism [3-4]. Such ST-driven magnetic precession is of fundamental interest for studying magnetic dynamics at the nanoscale and it can also be applied to make frequency-tunable microwave oscillators less than 100 nm in size. Conventionally, spin currents have been produced through the spin filtering effect of ferromagnetic electrodes in spin valves or magnetic tunnel junctions (MTJs), but recently it was demonstrated that a substantial spin current can alternatively be generated using certain non-magnetic materials through the spin Hall effect (SHE) [5-12]. By employing the SHE to inject spin current into an adjacent magnetic layer, it has been shown that one can induce magnetic switching in ferromagnetic (FM) metals [13-14] or excite dynamics in ferrimagnetic insulators with ultra low damping and small saturation magnetization [15]. Compared with conventional spin filtering, the optimized SHE can be more efficient in terms of applying a large ST to a FM layer [14] and it can also provide a versatile configuration for making highly reliable spin-torque devices [13].

In this work, we show that the SHE torque from direct current in a thin layer of $\beta$–Ta can excite persistent magnetic precession in a ferromagnetic metal. We detect the magnetic oscillations using a current-biased MTJ in contact with the oscillating element to produce a microwave output voltage across the MTJ. Our observation of magnetic oscillations in response to a direct current in the Ta layer provides additional confirmation for the ST origin of the current-induced switching we reported recently in similar devices [13]. Furthermore, we demonstrate that the three-terminal SHE-torque device geometry, by allowing separate variation of the ST driving current and the bias across the MTJ, enables independent control of the amplitude and frequency of ST nano-oscillators (STNOs).

We studied samples made from a Ta(6)/Co$_{40}$Fe$_{40}$B$_{20}$(1.5)/MgO(1.2)/Co$_{40}$Fe$_{40}$B$_{20}$(4)/Ta(5)/Ru(5) thin-film stack (thicknesses in nanometers), which was first patterned into a microstrip 1.2 μm wide and 6 μm long. The CoFeB/MgO/CoFeB MTJ was then milled into a 50 × 180 nm$^2$ nanopillar on the Ta microstrip . The thickness of the free CoFeB layer (1.5 nm) was chosen such that perpendicular interface anisotropy from Ta/CoFeB/MgO reduced the effective demagnetization field but left the easy magnetic axis in the film plane on average [16], thereby reducing the critical current required to excite ST-driven dynamics. We did not use annealing to enhance the tunneling magnetoresistance (TMR), as that would have resulted in a fully out-of-plane magnetic anisotropy for the free layer. We carried out all of our measurements at room temperature. The MTJ was oriented such that its long axis was perpendicular to the direction of the Ta strip (the direction of the current flow) and the external magnetic field was applied along this axis (Fig. 1(a)). The magnetic response of the MTJ near zero current is shown in Fig. 1(c), from which we can see that the free layer is saturated in-plane for moderate external magnetic fields $H_{app}$ and the applied field needed to cancel the dipole field from the fixed layer acting on the free layer is $H_{dip}$ = -80 Oe. The resistance of the Ta strip was measured independently to be ~ 3.6 kΩ. From the magnetic minor loop, the TMR of the MTJ can be determined to be ~ 18% under zero bias and the RA product of the MTJ is ~ 40 Ω-μm$^2$. The dependence of the MTJ resistance on its bias current for both the parallel (P) and antiparallel (AP) states is plotted in Fig. 1(d).

Our device geometry is chosen such that when a direct charge current is applied along the Ta strip the SHE injects spins into the CoFeB free layer that are either P or AP to the equilibrium position of the free layer magnetic moment, depending on the sign of the current (see Fig. 1(a)). For the case that the injected spin moments are AP to the free layer moment, they will exert a ST that reduces the effective magnetic damping. When the SHE spin current is large enough to reduce the net effective damping to zero, spontaneous steady-state magnetic precession will result. We used the circuit shown in Fig. 1(b) to excite and detect these SHE-torque driven magnetic dynamics. Two DC current sources with common ground were employed to separately apply current through the Ta strip and the MTJ. The current through the Ta strip $I_{Ta}$ injected a spin current into the free layer to excite magnetic dynamics, while the MTJ bias current $I_{MTJ}$ allowed electrical detection of the dynamics by converting the oscillations of the MTJ resistance $R_{rf}$ arising from the TMR into an oscillating voltage $V_{rf} = I_{MTJ}R_{rf}$. The output microwave power was amplified and then measured with a spectrum analyzer.

Spectra of the microwave power emitted by the device are shown in Fig. 2(a) for $I_{MTJ}$ = 60 µA and an applied external magnetic field $H_{app}$ = -160 Oe that initially aligns the free layer into the P state of the MTJ, and for different values of $I_{Ta}$ from -0.8 mA to 0.8 mA. (Here and in the rest of this paper the impedance mismatch between the device and the transmission cable has been taken into account; *i.e.*, the plotted powers correspond to values that would be emitted to a matched load.) Given our previous measurement that the ratio of the spin current density injected into the CoFeB relative to the charge current density in the Ta is $J_S/J_{Ta} \approx 0.15$ [13], we calculate that the spin current density applied by the SHE for $|I_{Ta}|$ = 0.8 mA should be $J_S \approx 1.7 \times 10^6$ $\hbar/(2e)$ A/cm$^2$. This is approximately ten times larger than the spin current density caused by the tunneling current, $J_{S,MTJ} \approx 1.9 \times 10^5$ $\hbar/(2e)$ A/cm$^2$ (for $I_{MTJ}$ = 60 µA and TMR = 18 %, Fig. 1(d)), so that one can expect the SHE torque to be dominant. Figure 2(a) shows that the magnetic dynamics are excited only for negative values of $I_{Ta}$. When $H_{app} > H_{dip}$, for which the free layer switches into the AP configuration of the MTJ, the asymmetry with respect to the sign of $I_{Ta}$ is reversed, with magnetic dynamics excited only for positive values of $I_{Ta}$ (not shown). This asymmetry is consistent with the switching behavior observed in Ref. [13], where it was observed that the SHE torque from negative (positive) currents destabilizes the P (AP) state. The fact that microwave signals are observed only for one sign of $I_{Ta}$ excludes the possibility that the observed phenomenon is merely thermal excitation of a ferromagnetic resonance mode.

When the magnetic moment of the free layer in an MTJ undergoes small angle precession at frequency $f_0$ about a collinear equilibrium position, the resistance should oscillate at $2f_0$ because $\Delta R \propto \cos\theta$ is an even function of the oscillation angle $\theta$. However, as shown in Fig. 2(a) we observe both a strong second harmonic peak and a weaker first harmonic peak. This suggests that there is a small misalignment between the equilibrium positions of the free layer and the fixed layer when $H_{app}= -160$ Oe. As is typical for STNOs, the lineshapes can be fitted reasonably well to Lorentzian peaks; for $I_{Ta}$ = -0.8 mA, $I_{MTJ}$ = 60 µA, and $H_{app}$ = -160 Oe we find (Fig. 2(b)) that the resonance frequency is $2f_0$ = 1.62 GHz and the full-width at half-maximum linewidth of the second harmonic peak is 104 MHz. The integrated power of the second harmonic signal is shown in Fig. 2(c) as a function of $I_{Ta}$, which indicates that the critical current to excite precession is approximately -0.2 mA. The dependence of the peak frequency $2f_0$ on $I_{Ta}$ is plotted in Fig. 2(d); as is typical for STNOs with in-plane anisotropy there is a substantial red shift of the peak as a function of increasing the absolute value of the drive current $I_{Ta}$, and hence increasing the oscillator energy. The dependence of the oscillation linewidth on $I_{Ta}$ is shown in the inset of Fig. 2(b).

In STNOs consisting of conventional 2-terminal spin valves or MTJs, the driving current to excite the dynamics and the sensing current to detect the dynamics are the same, but in our 3-terminal SHE device these two currents are separate, providing the opportunity to tune these two parameters independently. In Fig. 3(a), we show the microwave spectra obtained with different $I_{MTJ}$ while $I_{Ta}$ is held constant. Ideally, if the sensing current had no influence on the magnetic dynamics, one would expect that the only change in the output signal with $I_{MTJ}$ would be that output power $P$ should scale approximately as $I_{MTJ}^2$. In Fig. 3(b) we plot the integrated power $P$ (triangles) and the normalized power $P/\left[I_{MTJ}^2 T(I_{MTJ})\right]$ (circles) vs. $I_{MTJ}$ for the second harmonic peak, where $T(I_{MTJ}) = \Delta R\,(I_{MTJ})/\Delta R\,(I_{MTJ} = 0)$ accounts for the bias dependence of the DC magnetoresistance (inset of Fig. 1(d)). As expected, the normalized power is approximately constant, although there is a slow decrease as $I_{MTJ}$ increases from its greatest negative to its greatest positive value. We will comment further below on the origin of this slope.

A more dramatic effect of $I_{MTJ}$ on the magnetic oscillations is that the peak frequency undergoes a large blue shift with increasing $I_{MTJ}$, varying by 0.4 GHz (> 30%) over our bias range of ± 60 µA (Fig. 3(c)). Note that the frequency change is much larger for $\Delta I_{MTJ}$ = 120 µA than it is for $\Delta I_{Ta}$ = 0.6 mA in Fig. 2(d). Given that the spin current density associated with $I_{MTJ}$ is much smaller than the spin current density due to $I_{Ta}$ from the SHE (see above), the frequency shift is not reasonably attributable to the anti-damping ST of the tunneling current. The frequency shift with $I_{MTJ}$ furthermore has the wrong sign to be due to the ST of the tunneling current, and that mechanism would be inconsistent with the small change in the normalized microwave power in Fig. 3(b). Neither can any field-like torque or heating effect exerted by $I_{MTJ}$ explain this frequency shift, since both of these effects should be approximately even functions of $I_{MTJ}$ [17-18], in contrast to the approximately linear shift we observe. The substantial frequency shift as a function of $I_{MTJ}$ can however be related quantitatively to a change in perpendicular anisotropy of the free layer induced by variations in the electric field across the MgO tunnel barrier as $I_{MTJ}$ is varied [19-25]. Within a macrospin model, $f_0$ for small angle precession should obey the Kittel formula:

$$f_0 = (\gamma/2\pi)[(H_{ext} + H_c)(H_{ext} + H_{demag}^{eff} + H_c)]^{1/2} \approx (\gamma/2\pi)[H_{ext}(H_{ext} + H_{demag}^{eff})]^{1/2}. \tag{1}$$

Here $H_{ext} = H_{app} - H_{dip}$ is the net external field, $H_c$ is the within-plane anisotropy field (which is negligibly small in our samples) and $H_{demag}^{eff} = 4\pi M_S - 2K_u/M_S$ is the effective perpendicular demagnetization field where $M_S$ denotes the saturation magnetization and $K_u$ is the uniaxial anisotropy energy coefficient that can depend on the electric field across the tunnel barrier. Thus, a change in $K_u$ as a function of the bias across the MTJ will cause a shift in $f_0$.

To confirm that the frequency shift with $I_{MTJ}$ arises from changes in the free layer anisotropy, we measured the power spectra under different applied fields (Fig. 4(a)) with both applied currents kept constant at $I_{Ta} = -0.8$ mA and $I_{MTJ} = 60$ µA. The resultant peak frequencies are represented by the triangles in Fig. 4(b). (Note that the fundamental frequency $f_0$ is plotted here). Considering that there can be ~10% difference between the oscillation frequencies obtained under $I_{Ta} = -0.8$ mA and those corresponding to the onset current (Fig. 2(d)), we need to correct for the frequency shift caused by finite $I_{Ta}$. After taking this into account and fitting with Equation (1), we determined $H_{demag}^{eff}(I_{MTJ} = 60\ \mu A) = 1100 \pm 60$ Oe. This number is much smaller than the intrinsic demagnetization field for the CoFeB film ($4\pi M_S \approx 13{,}000$ Oe [13]), consistent with the large perpendicular anisotropy observed in the Ta/CoFeB/MgO system [16]. In Fig. 4(b), we also plot the measured oscillation frequency vs $H_{ext}$ for the case of $I_{MTJ} = -60$ µA, $I_{Ta} = -0.8$ mA (circles), for which the same fitting procedure determines $H_{demag}^{eff}(I_{MTJ} = -60\ \mu A) = 700 \pm 40$ Oe. Similar measurements at different values of $I_{MTJ}$ with fixed $I_{Ta} = -0.8$ mA yield the full dependence of $H_{demag}^{eff}$ on $V_{MTJ}$ shown in Fig. 4(c) ($V_{MTJ} = I_{MTJ}$ times the bias-dependent P state resistance of the MTJ, see Fig. 1(d)). We observe an approximately linear increase in $H_{demag}^{eff}$ when $V_{MTJ}$ is increased in the positive direction, with a mean slope of 730 ± 90 Oe/V. Taking into account the thickness of the free layer ($t_{free}$ = 1.5 nm), the thickness of the MgO barrier ($t_{MgO}$ = 1.2 nm), and $M_S$ = 1100 emu/cm$^3$, we determine that the rate of change of anisotropy energy versus electric field in our devices is $|d(K_u t_{free})/dE|$ = $[M_S t_{free} t_{MgO}/2] d(H_{demag}^{eff})/dV$ = 0.07 ± 0.01 erg/cm$^2$ (V/nm)$^{-1}$ or 70 ± 10 µJ/m$^2$ (V/nm)$^{-1}$. Here, in estimating the experimental uncertainty we have accounted for the effects of the weak ST due to spin filtering of $I_{MTJ}$ and due to a SHE produced by $I_{MTJ}$ after it leaves the MTJ and passes laterally through the Ta to ground. Our value for the voltage induced anisotropy change is consistent with previous reports, which range from 30 to 140 µJ/m$^2$ (V/nm)$^{-1}$ [19-25]. The sign of the effect is also consistent with previous experiments [21, 23-24]; that is, an electric field corresponding to positive values of $I_{MTJ}$ as is defined in Fig. 1(a) leads to an decrease in $K_u$ and hence a increase in $H_{demag}^{eff}$. The voltage-induced anisotropy change can furthermore account for the slope in normalized power versus $I_{MTJ}$ in our measurements (discussed above in relation to Fig. 3(b)), in that an increased $H_{demag}^{eff}$ will increase the critical current needed to excite the DC dynamics and hence decrease the precession angle for a given driving current, which results in a reduced microwave power.

Our 3-terminal SHE device provides an excellent platform for studying how ST and electric field-induced changes in magnetic anisotropy can separately influence magnetic dynamics in nanomagnets. Bias-dependent frequency shifts have been previously observed in the DC-current-induced magnetic dynamics of MTJs with thin free layers [2, 26] but it has been difficult to clearly differentiate between the effects of anisotropy changes and the ST. In contrast, in a 3-terminal SHE device one can employ MTJs with large resistance-area products for which only a very small tunneling current is needed to detect the free layer oscillation. This allows, as is illustrated above, separate analysis of the different physical mechanisms affecting the precessional dynamics.

In conclusion, we have demonstrated that the spin Hall effect from a direct current can be used to excite persistent magnetic oscillations in a metallic nanomagnet. We detect the oscillations using a MTJ in a 3-terminal device geometry. In comparison to conventional 2-terminal MTJ and spin valve STNOs, the 3-terminal SHE-based STNO allows separate control of the magnetic dynamics and the output electric power, therefore allowing greater and more versatile tuning of the frequency and amplitude of the microwave signal. The 3-terminal device also enables a quantitative measurement of the electric-field-induced magnetic anisotropy change at the CoFeB/MgO interface. The electric field control of the magnetic anisotropy of the free layer provides an efficient way to tune the oscillation frequency over a broad range, with over a 30% change demonstrated in this experiment.

This work was supported in part by the NSF/NSEC program through the Cornell Center for Nanoscale Systems, and by the Army Research Office and the Office of Naval Research. This work was performed in part at the Cornell NanoScale Facility, a node of the National Nanotechnology Infrastructure Network (NNIN) which is supported by the NSF (ECS-0335765), and benefited from use of the facilities of the Cornell Center for Materials Research, which is supported by the NSF/MRSEC program (DMR-1120296).

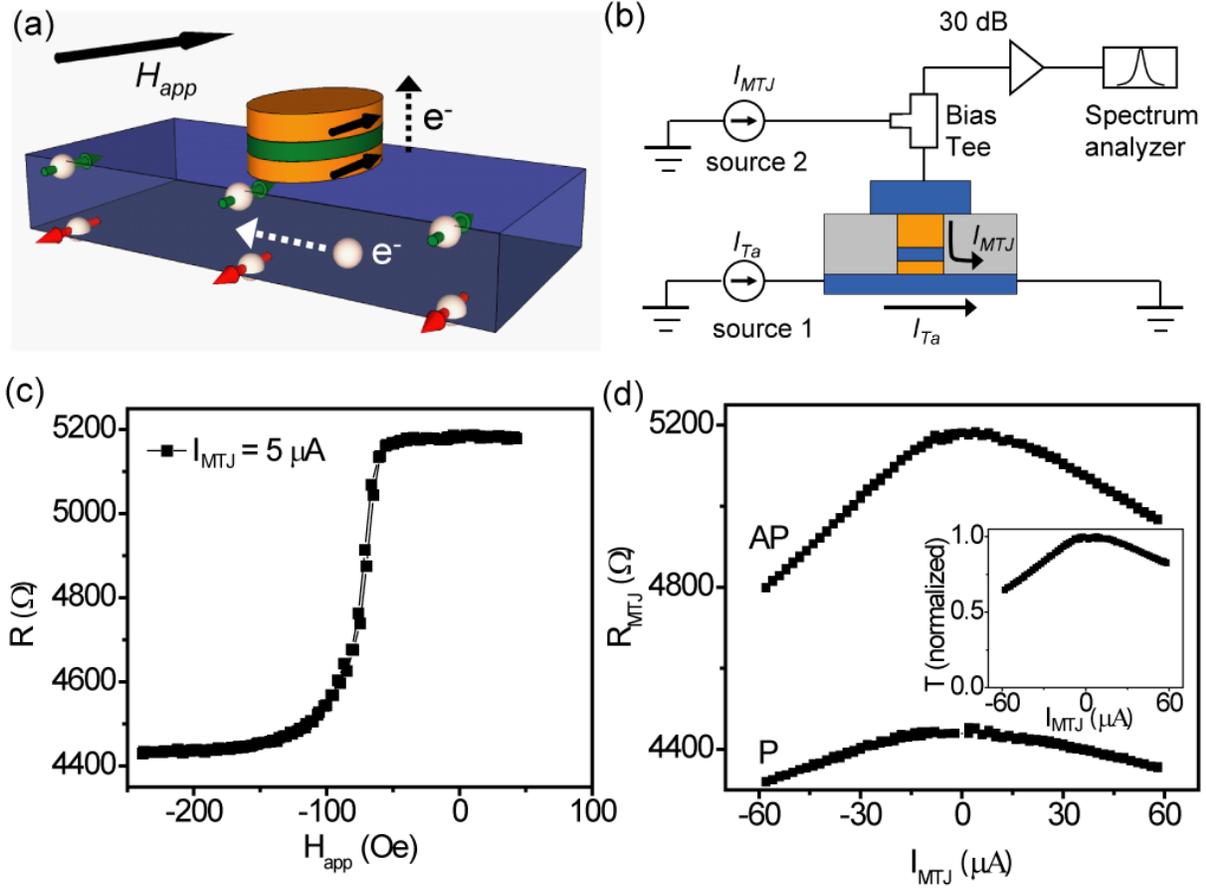

Figure 1. (a) Schematic showing the direction of spin accumulation induced by the SHE at the top and bottom surface of Ta strip by a positive current $I_{Ta}$. The dashed arrows denote the directions of electron flow for positive applied currents $I_{Ta}$ and $I_{MTJ}$. $H_{app}$ shown in the figure corresponds to a negative field, the direction used for our measurements of microwave spectra. (b) Circuit used to excite and detect the magnetic oscillations. (c) Magnetic minor loop of the MTJ. The lead resistance due to the Ta strip (0.5 × 3.6 kΩ = 1.8 kΩ) is already subtracted. (d) Bias current dependence of the MTJ DC resistance in the AP and P states. Inset to (d): Normalized magnetoresistance $T(I_{MTJ})$.

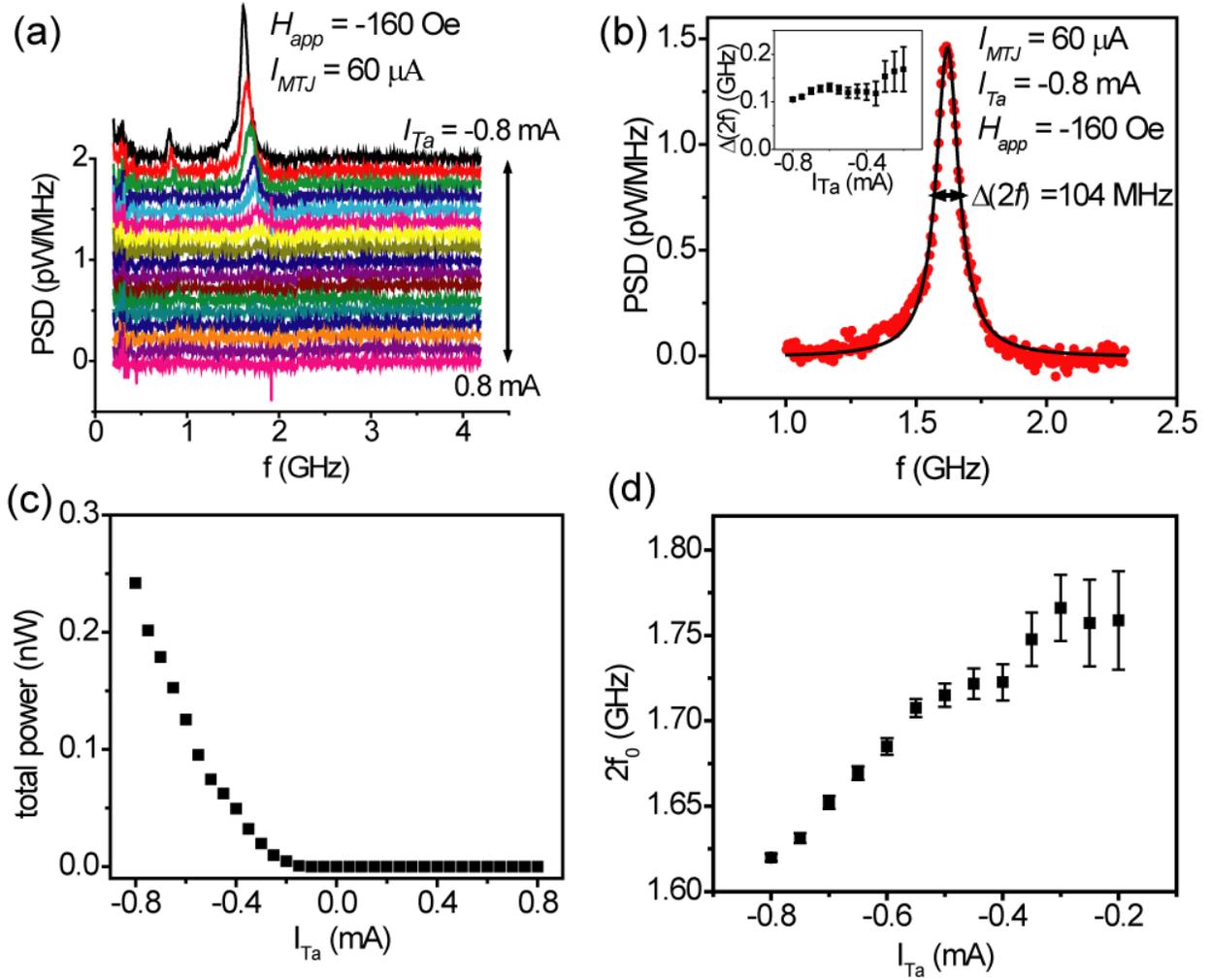

Figure 2. (a) Microwave power spectral densities measured for $H_{app}$ = -160 Oe, $I_{MTJ}$ = 60 μA and for $I_{Ta}$ ranging between -0.8 mA and +0.8 mA. The spectra under different currents are shifted vertically for comparison. (b) Microwave spectrum for $I_{Ta}$ = -0.8 mA in (a) with a fit to a Lorentzian peak. The linewidth of the second harmonic peak is 104 MHz. Inset: dependence of the linewidths for the second harmonic peaks as a function of $I_{Ta}$. (c) Total integrated power for the second harmonic peaks from (a). (d) Central frequency of the second harmonic peaks ($2f_0$) plotted versus $I_{Ta}$.

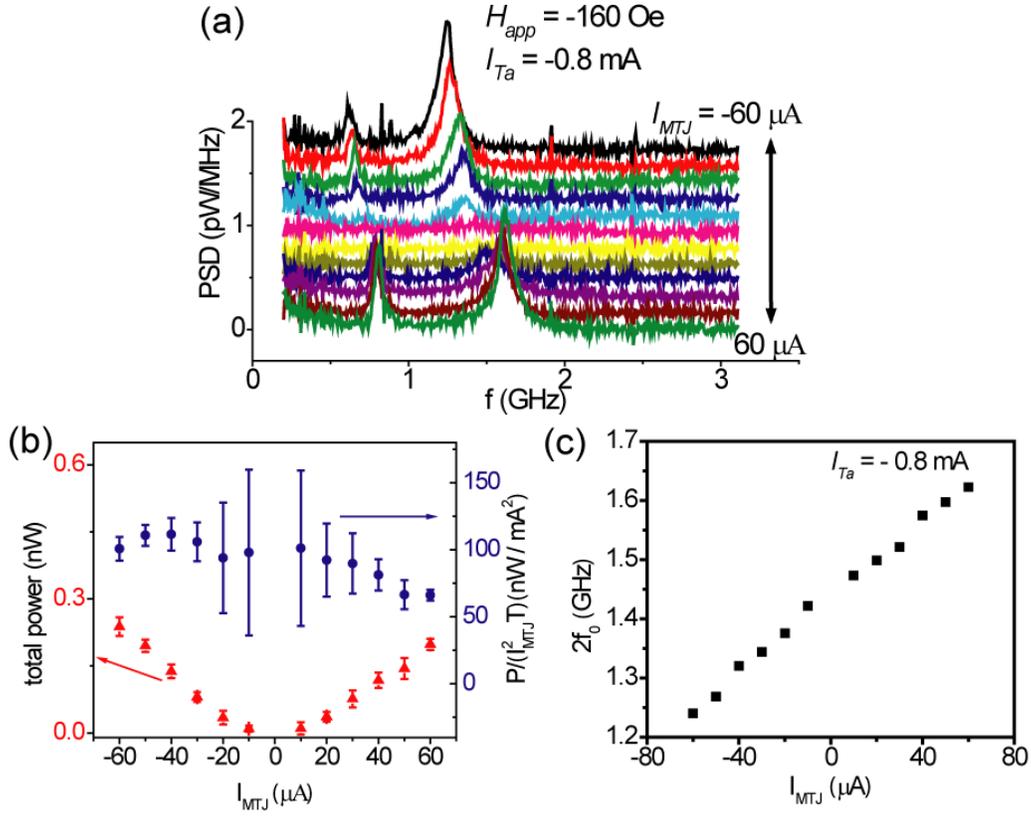

Figure 3. (a) Microwave spectra measured for $H_{app}$ = -160 Oe, $I_{Ta}$ = -0.8 mA and for $I_{MTJ}$ ranging between -60 μA and 60 μA. (b) Red triangles: integrated microwave power $P$ of the second harmonic peaks from (a) versus $I_{MTJ}$. Blue circles: integrated microwave power normalized by $I_{MTJ}^2$ and the magnetoresistance correction $T(I_{MTJ})$. (c) Central frequencies of the second harmonic peaks from (a) versus $I_{MTJ}$.

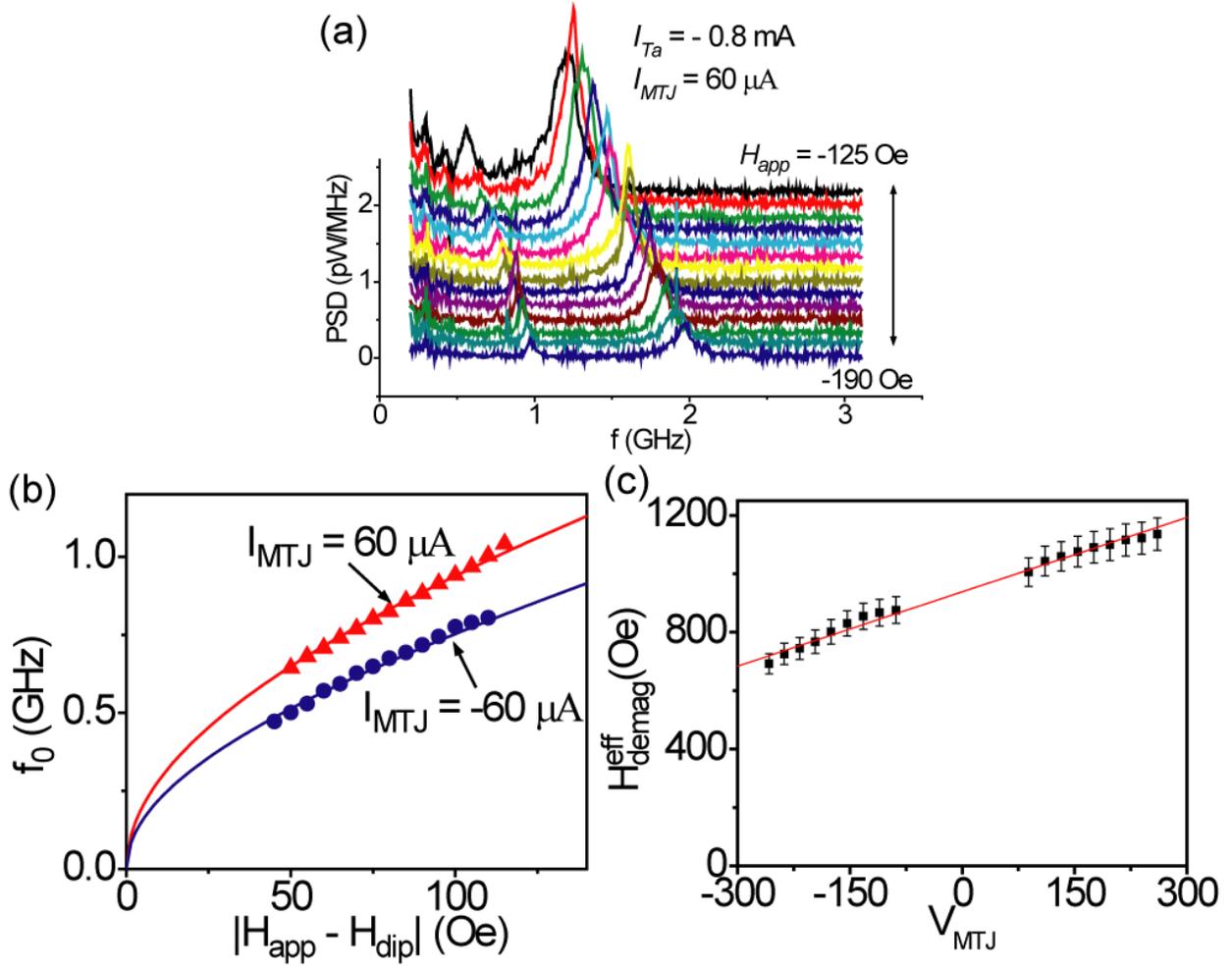

Figure 4. (a) Microwave spectra measured for $I_{Ta}$ = -0.8 mA, $I_{MTJ}$ = 60 μA and for $H_{app}$ ranging between -125 Oe and -190 Oe. (b) Measured oscillation frequencies $f_0$ plotted versus the net magnetic field acting on the free layer $|H_{app} - H_{dip}|$, where $H_{dip}$ = -80 Oe. The red triangles are obtained from the data in (a) and the blue circles are from spectra measured for $I_{Ta}$ = -0.8 mA and $I_{MTJ}$ = -60 μA. The solid lines represent fits to the Kittel formula. (c) Dependence of the effective demagnetization field $H_{demag}^{eff}$ on $V_{MTJ}$, with a linear fit.